\documentclass[eqsecnum,floats,aps,nofootinbib,showpacs]{revtex4}

\renewcommand{\d}{\mathrm{d}}
\renewcommand{\l}{\left(}
\renewcommand{\r}{\right)}

\usepackage{amsmath, amsthm, amssymb}

\def\be{\begin{equation}}
\def\ee{\end{equation}}
\def\beq{\begin{equation*}}
\def\eeq{\end{equation*}}
\def\ba{\begin{aligned}}
\def\ea{\end{aligned}}

\def\ov{\overline}

\begin{document}
\title{The Poincar\'e Gauge Theory of Gravty and the Immirzi Parameter}

\author {Marcin Ka\'zmierczak}
\email{marcin.kazmierczak@fuw.edu.pl}
\affiliation{Institute of Theoretical Physics, Uniwersytet Warszawski, Ho\.{z}a 69, 00-681 Warszawa, Poland} 

\begin{abstract}
The minimal coupling method proved to yield definite and correct physical predictions when applied to fundamental fermions within the framework of Yang--Mills theories of Standard Model. Similarly, the possibility of formulating gravity as the Poincar\'e gauge theory gives the opportunity to produce definite predictions for fermions in the presence of gravitational field. The minimal coupling procedure, however, cannot be applied naively but rather needs to be modified slightly such that it is unambiguous. Application of the corrected coupling method to fermions, together with the inclusion of the Holst term in the gravitational part of the action, leads to the conclusion that the Immirzi parameter is in principle classically measurable, in agreement with the result of Perez and Rovelli.
\end{abstract}

\pacs{04.50.Kd, 04.60.Bc, 04.60.Ds, 04.60.Pp, 04.80.Cc}

\maketitle
\section{Introduction}\label{section1}

All the relativistic theories in the absence of
gravity are invariant under the (global) action of the Poincar\'e group. As shown by by Kibble
\cite{Kib1}, it is possible to obtain a reasonable theory of gravity by simply localizing this global symmetry.  This approach necessitates first order
formalism for general relativity, with metric (but non--symmetric)
connection, as the set of Yang--Mills fields has to consist of ten
independent one--forms. Interestingly, this formalism appears to simplify the canonical analysis and quantization of the theory and therefore it is employed in Loop Quantum Gravity (LQG), although in standard LQG the time gauge is imposed at the very beginning that breaks the gauge group effectively to $SU(2)$.

The
natural way to introduce the interaction within the
spirit of Yang--Mills theory is to use the minimal coupling
procedure (MCP). Indeed, in the standard model of particle physics
this procedure is followed on the fundamental level, leading to predictions that agree with experimental results with great
accuracy. 
Also, in GR the principle of equivalence, which states that the
effects of gravitation can be locally `turned off' by a suitable
choice of a reference frame, necessitates minimal coupling \cite{WeinGr}. However, in the Poincar\'e gauge theory of gravity MCP appears to be ambiguous. The ambiguity
 is of importance for the standard EC theory with fermions \cite{Kazm1}, as well as the theory modified by the addition of the Holst term \cite{Kazm2}. The predictions concerning fermions in the Ashtekar--Barbero--Immirzi formalism obtained in \cite{PR} can be radically changed if the freedom of adding divergence of a vector field to the initial fermionic Lagrangian density is exploited \cite{Kazm2}. 

Luckily, the corrected unambigoues coupling procedure has been proposed \cite{Kazm5} which makes the predictions of the theory unique. They appear to agree with those derived by Perez and Rovelli.

\section{The Poincar{\'e} gauge theory}\label{section2}
\subsection{Yang--Mills theories}\label{YM}

The leading idea of standard Yang--Mills theories is that any interaction can be associated with a symmetry group. 
Let$G$ be a Lie group and let
\be\label{S}
S_m[\phi]=\int\mathcal{L}_m\l{\phi,\partial_{\mu}\phi}\r
d^4x=\int\mathfrak{L}_m\l{\phi,d\phi}\r
\ee
represent the action of a field theory of a matter field $\phi$ in Minkowski space $M$. Here
$\mathcal{L}_m$ is a Lagrangian density and $\mathfrak{L}_m$ a Lagrangian four--form. Assume
that $\mathcal{V}$ is a (finite dimensional) linear space in which fields $\phi$ take their values,
$\phi:M\rightarrow \mathcal{V}$, and $\pi$ is a representation of $Lie(G)$ on $\mathcal{V}$.
Let $\rho$ denote the corresponding representation of the
universal covering group of $G$, 
$\rho\l{\exp (\mathfrak{g})}\r=\exp\l{\pi(\mathfrak{g})}\r$. Suppose
that the Lagrangian four--form, and hence the action, is invariant under its global action
\be\label{global}
\mathfrak{L}_m\l{\rho(g)\phi,d\l{\rho(g)\phi}\r}\r=\mathfrak{L}_m\l{\rho(g)\phi,\rho(g)d\phi}\r=
\mathfrak{L}_m\l{\phi,d\phi}\r .
\ee
Then one can introduce an interaction associated to the symmetry group
$G$ by allowing the group element $g$ to depend on space--time point
and demanding the theory to be invariant under the local (space--time point dependent) action of
$G$. This can be most easily achieved by replacing the
differential by the covariant differential
\be\label{MCPYM}
d\phi\rightarrow D\phi=d\phi+\mathbb{A}\phi ,
\ee
where $\mathbb{A}$ is a $Lie(G)$--valued\footnote{More precisely,
  $\mathbb{A}$ and $\mathbb{F}$ take values in the representation
  $\pi$ of $Lie(G)$. The same remark applies to the analogues situations
  later on.}  one--form field on $M$ which
transforms under the local action of $G$ as
\be\label{gauge}
\mathbb{A}\rightarrow\mathbb{A}'=\rho(g)\mathbb{A}\rho^{-1}(g)-d\rho(g)\rho^{-1}(g) .
\ee
Then $D'\phi'=\rho(g)D\phi$ and as a result the Lagrangian that includes the interaction, defined by 
\be\label{MCP}
\tilde{\mathfrak{L}}_m\l{\phi,\d\phi,\mathbb{A}}\r:=\mathfrak{L}_m\l{\phi,D\phi}\r ,
\ee
is invariant under the local action of $G$, on account of (\ref{global}). The method (\ref{MCP}) of constructing $\tilde{\mathfrak{L}}_m$ from $\mathfrak{L}_m$ is referred to as {\it minimal coupling procedure} (MCP). 
The next step is the construction of a gauge--field part
of the Lagrangian $\mathfrak{L}_G$, which should be built of $\mathbb{A}$ and remain
invariant under gauge transformations (\ref{gauge}). The total Lagrangian four--form of the theory is then
\be
\mathfrak{L}=\tilde{\mathfrak{L}}_m+\mathfrak{L}_G .
\ee

\subsection{The Poincar{\'e} group as a gauge group}\label{PGGG}
Let us inquire the properties of an interaction associated to the Poincar\'e group $\mathcal{P}$.
Let 
\beq
\ba
&\rho(\Lambda,a):=\rho(a)\rho(\Lambda),\\
&\rho(a):=\exp\l{a_{a}P^a}\r,\quad 
\rho\l{\Lambda(\varepsilon)}\r:=\exp\l{\frac{1}{2}\varepsilon_{ab}J^{ab}}\r
\ea
\eeq
be the representation of $\mathcal{P}$. Here $P^a, J^{ab}$ are the
generators of translations and Lorentz rotations and belong to the
representation $\pi$ of $Lie(\mathcal{P})$. The coefficients $\varepsilon_{ab}=-\varepsilon_{ba}$ are
parameters of the Lorentz transformation 
$\Lambda(\varepsilon)=\exp (\varepsilon)$. Here $\varepsilon\in so(1,3)$
is the matrix with entries
${\varepsilon^a}_b:=\eta^{ac}\varepsilon_{cb}$, where
$(\eta^{ab})=(\eta_{ab})=diag(1,-1,-1,-1)$ is Minkowski matrix. Using
the composition law and employing infinitesimal transformations one
can derive transformation properties of the generators for the
Poincar{\'e} algebra
\be\label{transcom}
\ba
&\rho(\Lambda,a)P^a\rho^{-1}(\Lambda,a)={\Lambda_c}^aP^c,\\ 
&\rho(\Lambda,a)J^{ab}\rho^{-1}(\Lambda,a)={\Lambda_c}^a{\Lambda_d}^b\l{J^{cd}+a^cP^d-a^dP^c}\r .
\ea
\ee
In order to construct the covariant
differential, one needs to introduce the $Lie(\mathcal{P})$-valued one--form
\be\label{pA}
\mathbb{A}=\frac{1}{2}\omega_{ab}J^{ab}+\Gamma_aP^a ,
\ee
where $\omega_{ab}=-\omega_{ba}$ and $\Gamma_a$ are
one--forms\footnote{Here $\mathcal{M}$ is the space--time manifold,
  which will no longer be the Minkowski space $M$ in the presence of gravity.} on
$\mathcal{M}$. Under gauge transformations (\ref{gauge}), these one--forms
transform as
\be\label{gftrans}
\omega'=\Lambda\omega\Lambda^{-1}-d\Lambda\Lambda^{-1}, \quad
\Gamma'=\Lambda\Gamma-\omega'a-da ,
\ee
which can be easily verified for infinitesimal transformations (use
(\ref{transcom})). Here $\omega$ is a matrix with entries
${\omega^a}_b$ and $\Gamma$ a column matrix with entries
$\Gamma^a$. The first
equation of (\ref{gftrans}) looks like the transformation rule for
connection one--forms under the change of an orthonormal
frame of vector fields. Indeed, orthonormal frames on Lorentzian manifold are connected
by local Lorentz transformations. Note also that the antisymmetry
$\omega_{ab}=-\omega_{ba}$ means metricity of the resulting
space--time connection. The metric on space--time is obtained from a cotetrad
field $e$ that needs to be somehow constructed from translational gauge fields. The transformation
formula for $e$, compatible with the one for $\omega$ (\ref{gftrans}),
would be $e'=\Lambda e$, so one cannot just adopt the
translational gauge field $\Gamma$ as representing cotetrad. The
solution is to introduce a vector--valued zero--form $y$ (a column of
functions on $\mathcal{M}$), which transforms under the local
Poincar{\'e} transformation $(\Lambda,a)$ according to
\be\label{ytrans}
y'=\Lambda y+a ,
\ee
 and then introduce the cotetrad
\be\label{e}
e:=\Gamma+dy+\omega y .
\ee
Then from (\ref{gftrans}) and (\ref{ytrans}) it follows that
$e'=\Gamma'+dy'+\omega'y'=\Lambda e$, as desired. The new
field $y$ can be given a natural geometric interpretation in the language of fibre
bundles \cite{T1}, as well as the physical meaning
\cite{GN}\cite{T3h}\cite{Lh}. What is more, if the Lagrangian
four--form depends on $y$ and $\Gamma$ only via the cotetrad $e$, one
is free to acknowledge $e$ as a fundamental field and forget about its
origin. Indeed, the variation of such a Lagrangian would be
\be\label{ew}
\delta{\mathfrak{L}}_m=
\delta e^a\wedge\frac{\delta\mathfrak{L}_m}{\delta e^a}+
\delta\omega^{ab}\wedge\frac{\delta\mathfrak{L}_m}{\delta\omega^{ab}}+
\delta\phi\wedge\frac{\delta\mathfrak{L}_m}{\delta\phi}  ,
\ee
$\phi$ representing matter fields. Since $\delta
e^a=\delta\Gamma^a+D\delta y^a+\delta{\omega^a}_by^b$, we finally get
\be\label{gyw}
\ba
&\delta{\mathfrak{L}_m}=
\delta \Gamma^a\wedge\frac{\delta\mathfrak{L}_m}{\delta e^a}-
\delta y^a D\l{\frac{\delta\mathfrak{L}_m}{\delta e^a}}\r\\
+&\delta\omega^{ab}\wedge\l{y_b\frac{\delta\mathfrak{L}_m}{\delta e^a}+\frac{\delta\mathfrak{L}_m}{\delta\omega^{ab}}}\r+
\delta\phi\wedge\frac{\delta\mathfrak{L}_m}{\delta\phi}+d\l{\delta y^a\frac{\delta\mathfrak{L}_m}{\delta e^a}}\r .
\ea
\ee
Comparing (\ref{gyw}) with (\ref{ew}) one can see that promoting $e$
to the fundamental field, instead of $\Gamma$ and $y$, do not
influence the resulting system of field equations.

Using $\omega$ and $e$ we define the {\it torsion two--form} $Q$ and the {\it curvature two--form} $\Omega$ as
\be
Q^a:=De^a=\frac{1}{2}{T^a}_{bc}e^b\wedge e^c ,\qquad
{\Omega^a}_b:=d{\omega^a}_b+{\omega^a}_c\wedge{\omega^c}_b=\frac{1}{2}{R^a}_{bcd}e^c\wedge e^d .
\ee
The simplest possible Lagrangian that is constructed from $\omega$ and $e$ in the Poincar\'e invariant way is the Palatini one, $\mathfrak{L}_G=-\frac{1}{4k}\epsilon_{abcd}e^a\wedge e^b\wedge
\Omega^{cd}$. However, it is also alowable to add the Holst term and adopt
\be
\mathfrak{L}_G=-\frac{1}{4k}\epsilon_{abcd}e^a\wedge e^b\wedge \Omega^{cd}+
\frac{1}{2k\beta}e^a\wedge e^b\wedge \Omega_{ab} 
\ee
as a gauge field part of the Lagrangian, since the Holst term is also invariant under the local Poincar\'e transformations. \footnote{Note, however, that the behavior of $\mathfrak{L}_{hol}$ under parity is different from that of Palatini term. Hence, if parity is thought of as a fundamental symmetry of the theory, then the Holst term should not be added.} The parameter $\beta$ representing the relative weight of the two terms is called the {\it Immirzi parameter}.

\section{The ambiguity of MCP in the presence of torsion}

It is well known that the transformation of divergence addition to the matter lagrangian density in Minkowski space,
\beq\label{Lch}
\mathcal{L}_m\rightarrow\mathcal{L}_m'=\mathcal{L}_m+\partial_{\mu}V^{\mu},
\eeq
is a symmetry of the theory, at least so long as global topological aspects are not taken into account (not only field equations remain unchanged, but also Noether conserved currents of physical interest preserve their form under (\ref{Lch})).
When choosing the method of obtaining
$\tilde{\mathfrak{L}}_m\l{\phi,d\phi,\mathbb{A}}\r$ from $\mathfrak{L}_m\l{\phi,d\phi}\r$,
it is necessary to impose a consistency requirement that the transformation (\ref{Lch}) is a
symmetry of the resulting final theory, in with the interaction is turned on. In the case of Poincar\'e gauge theory, it appears that 
MCP, defined by (\ref{MCP}), does not satisfy this requirement! If MCP is applied to the differential
$\partial_{\mu}V^{\mu}d^4x=d\l{V\lrcorner d^4 x}\r$, it will pass to
$d\l{V\lrcorner\epsilon}\r-T_aV^a\epsilon$. This expression is not a topological term, unless the torsion trace vanishes.
Now, in the case of the Dirac field, 
there exists a two--parameter family of equaly good flat space Lagrangians given by
\be\label{LF}
\mathcal{L}_F=\mathcal{L}_{FR}+\partial_{\mu}V^{\mu},\qquad V^{\mu}=aJ_{(V)}^{\mu}+bJ_{(A)}^{\mu}, \qquad a,b\in\mathbb{R},
\ee
where $J_{(V)}^{\mu}=\ov{\psi}\gamma^{\mu}\psi$ and $J_{(A)}^{\mu}=\ov{\psi}\gamma^{\mu}\gamma^5\psi$ are the Dirac vector and axial currents.
A shown in  \cite{Kazm1} and  \cite{Kazm1}, drastically different prediction of both EC and Holst--modified EC theory can be obtained by changing the parameters $a$ and $b$ of (\ref{LF}) if MCP is used.

For the sake of comparison of the theory with standard general relativity, it is usefull to use the algebraic relation between the spin density and the torsion to express the latter through matter fields. The result can then be inserted back to the initial Lagrangian. In this way, an effective Lagrangian is obtained that does not depend on torsion anymore. It can be used to derive the equations relating the space--time metric with matter fields. If applied to $\mathfrak{L}=\tilde{\mathfrak{L}}_F+\mathfrak{L}_G$, where $\tilde{\mathfrak{L}}_F$ is a result of application of MCP to $\mathfrak{L}_F=\mathcal{L}_F d^4x$ deffined in (\ref{LF}), the resulting effective Lagrangian is given, up to the total differential, by
\be
\ba
&\mathfrak{L}_{eff}={\stackrel{\circ}{\mathfrak{L}}_G}+\stackrel{\circ}{\tilde{\mathfrak{L}}}_{FR}+\l{C_{AA}
    \,J^{(A)}_aJ_{(A)}^a+C_{AV} \,J^{(A)}_aJ_{(V)}^a+C_{VV}
    \,J^{(V)}_aJ_{(V)}^a}\r\epsilon , \\
&C_{AA}=\frac{3k\beta}{16(1+\beta^2)}\left[{4b+\beta(1-4b^2)}\right], \quad 
C_{AV}=\frac{3k\beta}{4(1+\beta^2)}a(1-2\beta b), \quad
C_{VV}=-\frac{3k\beta^2}{4(1+\beta^2)}a^2.
\ea
\ee
Here a circle above an object means that a Levi--Civita connection is used in them.
The $C$s clearly play a role of coupling constants preceding the gravity--induced four--fermion point interaction terms which would not be present in standard general relativity. In the special case of $a=b=0$ one obtains the result discussed by Perez and Rovelli \cite{PR}. These authors concluded that the Holst--modified Einstein--Cartan theory with fermions differs from general relativity by the presence of point fermion interaction of axial--axial type with very small coupling constant that depends on the Immirzi parameter. However, chosing $a=b=0$ is just as equally good as chosing $a=1$ and $b=3$, or any other choice. Therefore, the discussion performed by Perez and Rovelli was not really conclusive. In order to have a definite result and conclude whether the Immirzi parameter is classically measurable, it is first necessary to replace MCP by a corrected, unambiguous coupling method.

\section{Removing the ambiguity by modifying the coupling procedure}

Let us again consider the action (\ref{S}) for a matter field $\phi$, which takes values in a linear space $\mathcal{V}$ and exhibits invariance under global action of a Lie group $G$ according to (\ref{global}). In order to localize the symmetry, perform the replacement  
\be
d\phi\rightarrow \mathcal{D}\phi=d\phi+\mathcal{A}\phi,
\ee
where $\mathcal{A}$ is a $Lin(\mathcal{V})$--valued one--form field on
$M$ satisfying
\be\label{Acaltr}
\mathcal{A}\rightarrow\mathcal{A}'=\rho(g)\mathcal{A}\rho^{-1}(g)-d\rho(g)\rho^{-1}(g) .
\ee
Note the similarity of this construction with (\ref{MCPYM}). There is, however, a crucial difference in that $\mathcal{A}$ is not required to take values in the range of the representation $\pi$ of $Lie(G)$ but in the larger set of all linear maps of $\mathcal{V}$ into itself. The transformation formula (\ref{Acaltr}) alone suffices to guarantee local invariance of the final action. 

If there exists a natural $\rho$--invariant scalar product structure on $Lin(\mathcal{V})$, which is usually the case \cite{Kazm5}, then  
$\mathcal{A}$ can be decomposed as
\be
\mathcal{A}=\mathbb{A}+\mathbb{B}(\mathbb{A},e) ,
\ee
where $\mathbb{A}$ is $Ran(\pi)$--valued and $\mathbb{B}(\mathbb{A},e)$ is $Ran(\pi)^{\perp}$--valued. Here ${}^{\perp}$ denotes 
an orthogonal complement (see \cite{Kazm5} for the definition of the appropriate scalar product for the Dirac field). Note that $\mathbb{B}$ is that part of $\mathcal{A}$ that corresponds to the deviation of this procedure from standard YM. It is required to be determined by the Yang--Mills field $\mathbb{A}$ and, in the case of gravity, the cotetrad (recall that the cotetrad is composed of Yang--Mills fields and the Poincar\'e coordinates).

It appears that the requirement that the construction be free of the ambiguity, together with some other natural assumptions which may be specific for the particular matter field, fixes the form of $\mathbb{B}$ uniquely. Specifically, in the Dirac field case, the requirements are that\newline
1) $\mathbb{B}$ transforms appropriately under gauge transformations
\be
\mathbb{B}(\mathbb{A}',e')=\rho(g)\mathbb{B}(\mathbb{A},e)\rho^{-1}(g) ,
\ee
2) the the procedure is free of the ambiguity ,\newline
3) the Leibniz rule holds
\be
\ba
&(\mathcal{D}\ov{\psi})\gamma^a\psi+\ov{\psi}\gamma^a\mathcal{D}\psi=
dJ_{(V)}^a+\tilde{\omega}{^a}_bJ_{(V)}^b,\\
&(\mathcal{D}\ov{\psi})\gamma^a\gamma^5\psi+\ov{\psi}\gamma^a\gamma^5\mathcal{D}\psi=
dJ_{(A)}^a+\tilde{\omega}{^a}_bJ_{(A)}^b,
\ea
\ee
where $\mathcal{D}\ov{\psi}:=(\mathcal{D}\psi)^{\dag}\gamma^0$ and
$\tilde{\omega}{^a}_b$ is an auxiliary connection on space--time, which may differ from $\omega{^a}_b$.

As shown in \cite{Kazm5}, the solution of these conditions is provided by
\be
\ba
&\tilde{\omega}{^a}_b={\omega^a}_b+\mathbb{T}\delta^a_b ,\\
&\mathcal{D}\psi=d\psi-\frac{i}{4}\omega_{ab}\Sigma^{ab}+\frac{1}{2}\mathbb{T}{\bf 1}+i\mu_1{\bf 1}+i\mu_2\gamma^5 ,
\ea
\ee
where $\mu_1, \mu_2$ are scalar real--valued one--forms and $\mathbb{T}=T_ae^a$ is the torsion--trace--one--form. Note that the one--forms $\mu$ and $\nu$ do not influence the form of the modified space--time connection $\tilde{\omega}$.
If other fundamental interactions were also considered, then these one--forms could be hidden in the gauge fields corresponding to the $U(1)$ symmetry and the approximate chiral symmetry of weak interactions. Therefore it is reasonable to set them to zero when considering gravity.

The resulting effective action for the Einstein--Cartan gravity with Holst term and fermionic matter turns out to be equal to
\be
\mathfrak{L}_{eff}={\stackrel{\circ}{\mathfrak{L}}_G}+\stackrel{\circ}{\tilde{\mathfrak{L}}}_{FR}+
\frac{3k\beta^2}{16(1+\beta^2)}\,J^{(A)}_aJ_{(A)}^a\,\epsilon ,
\ee
which is identical with the result of \cite{PR}.

\section{conclusions}

The modified coupling procedure provides a consistent method for coupling gravity to other field theories within the framework of the Poincar\'e gauge theory of gravity. The results do not involve additional parameters and do not depend on adding topological terms to the initial action. For standard gauge theories, the procedure reduces to MCP (this was not shown here, but the verification is straightforward).

The advantage of the application of the new method of coupling in the discussion of the Holst--modified Einstein--Cartan theory with fermions is merely to re-derive the result of \cite{PR} in an unambiguous way. The final conclusion is that the theory differs from general relativity with fermions by the presence of axial--axial four--fermion point interaction whose coupling constant depends on the Immirzi parameter. Hence, if the strength of such interaction was measured, an information about the value of $\beta$ would be obtained. If compared with the value of $\beta$ from black hole entropy, such measurement could provide an experimental test of LQG, since it is reasonable to expect that the classical limit of this theory should be compatible with the initial classical theory that was quantized (otherwise the very quantization procedure would fail to pass the basic test of consistency). Of course, any disagreement could be due to the incorrectness of the Beckenstein--Hawking formula and not necessarily LQG itself. 

Unfortunately, careful dimensional analysis shows that the value of this coupling constant is of order of the square of the Planck length. Therefore, there is no hope of performing that kind of measurement in  reasonable future.

\end{document}